\documentclass[iop]{emulateapj}

\shorttitle{Orthogonal GRB} \shortauthors{Contopoulos et al.}

\def\gsim{\mathrel{\raise.5ex\hbox{$>$}\mkern-14mu
             \lower0.6ex\hbox{$\sim$}}}
\def\lsim{\mathrel{\raise.3ex\hbox{$<$}\mkern-14mu
             \lower0.6ex\hbox{$\sim$}}}

\begin{document}

\author{Ioannis Contopoulos\altaffilmark{*},
Antonios Nathanail\altaffilmark{1} and Daniela Pugliese}
\affil{Research Center for Astronomy, Academy of Athens, Athens
11527, Greece}

\title{The orthogonal gamma-ray burst model}

\altaffiltext{1}{Also at Department of Physics, University of
Athens, Athens 15783,
Greece}
\altaffiltext{*}{icontop@academyofathens.gr}


\begin{abstract}We explore the analogy between a rotating
magnetized black hole and an axisymmetric pulsar and derive its
electromagnetic spindown after its formation in the core collapse
of a supermassive star. The spindown shows two characteristic
phases, an early Blandford-Znajek phase that lasts a few hundred
seconds, and a late pulsar-like afterglow phase that lasts much
longer. During the first phase, the spindown luminosity decreases
almost exponentially, whereas during the afterglow phase it
decreases as $t^{-a}$ with $1\lsim a \lsim 1.5$. We associate our
findings with long duration gamma-ray bursts (GRB) and compare
with observations.
\end{abstract}

\keywords{Pulsars; black holes; gamma-ray bursts; magnetic fields}


\section{Neutron star electrodynamics}

We explore black hole electrodynamics through the analogy with the
axisymmetric pulsar (Goldreich \& Julian~1969; Contopoulos,
Kazanas \& Fendt~1999). We begin with three simple statements:

1) {\em A neutron star may be charged}. One might naively argue
that once you embed a charged star in an ionized medium it will
attract carriers of the opposite charge and very quickly will lose
its charge. A relativistic astrophysicist, however, will argue
that a neutron star with a dipole magnetic field spinning along
its magnetic axis inside an ionized medium (not vacuum), induces a
distribution of radial electric field
\begin{equation}
E_r=B_\theta\frac{\Omega r_*\sin\theta}{c}=B \frac{\Omega
r_*}{c}\sin^2\theta \label{Er}
\end{equation}
(Goldreich \& Julian~1969), and therefore an electric charge
\begin{equation}
Q=\int_0^{\pi}2\pi r_*^2 \sin\theta E_r {\rm
d}\theta=\frac{8\pi}{3}r_*^2 B \frac{\Omega r_*}{c}\ .\label{Q}
\end{equation}
Here, $B$ is the equatorial value of the dipole magnetic field as
measured by a non-rotating observer, $\Omega$ and $r_*$ are the
angular velocity and the radius of the star, and $\theta$ the
polar angle. This charge is distributed in the neutron star
interior in such a way so as to satisfy the infinite conductivity
condition $E\cdot B=0$ everywhere. It may be sitting inside an
ionized magnetosphere, but it is not `sitting idle' waiting to be
discharged. The spinning neutron star is an astrophysical engine
that electrically polarizes its magnetosphere, generates large
scale electric currents, and emits electromagnetic (Poynting)
radiation. As long as this engine operates, the neutron star is
not discharged. We will see in the next section that a similar
result may also apply to black holes.

2) {\em A neutron star supports its own magnetic field}. What is
of interest here is that an observer co-rotating with the neutron
star measures an intrinsic dipole magnetic field $B_*$ generated
by toroidal electric currents in the neutron star interior.
However, the magnetic field $B$ measured by a stationary observer
is different. $B$ is the Lorentz transformation of $B_*$, namely
\begin{equation}
B=\frac{B_*}{\left[1-(\Omega r\sin\theta/c)^2\right]^{1/2}}\approx
B_*+\frac{1}{2}\left(\frac{\Omega r\sin\theta}{c}\right)^2 B_*\ ,
\label{deltaB1}
\end{equation}
with $\Omega r_*/c$ typically less than about 0.1. An equivalent
way to view this result is that the intrinsic stellar magnetic
field induces a certain distribution of charge in the stellar
interior and in the rotating magnetosphere, and thus forms a
distribution of toroidal currents that generates an extra poloidal
magnetic field component
\begin{equation}
\delta B \sim \frac{Q}{r_*^2}\left(\frac{\Omega r_*}{c}\right)\ .
\label{deltaB1}
\end{equation}
As we will see below, a similar result may also apply to black
holes.

3) {\em Isolated neutron stars spin down electrodynamically}. We
remind the reader that the magnetosphere of the axisymmetric
pulsar consists of closed and open field lines, and only the open
field lines (those that cross the light cylinder) contribute to
the neutron star spindown as
\begin{equation}
\dot{E} =\frac{2}{5}M_* r_*^2 \Omega\dot{\Omega}= -\frac{\Psi_{\rm
open}^2\Omega^2}{6\pi^2 c} = -B^2 r_*^2 c \left(\frac{\Omega
r_*}{c}\right)^4  \label{EdotAP}
\end{equation}
(Contopoulos~2005). Here, $M_*$ is the mass of the neutron star.
One can solve eq.~(\ref{EdotAP}) to obtain $\Omega=\Omega(t)$ and
$\dot{E}=\dot{E}(t)$, and thus easily show that at late times,
\begin{eqnarray}
\Omega & \propto & t^{-1/2}\ ,\ \ \mbox{and}\label{Omegadotp}\\
\dot{E}& \propto & t^{-2}\ .\label{Edotp}
\end{eqnarray}
We will now see that, under certain astrophysical circumstances,
rotating black holes may too function as axisymmetric pulsars.

\section{Black hole electrodynamics}

We will consider black holes that form in the core collapse of
supermassive stars. If the star is magnetized, magnetic flux will
be advected with the collapse. The material that is going to
collapse into a black hole will be strongly magnetized, and
therefore its core will pass through a spinning magnetized neutron
star stage. A certain amount of magnetic flux and electric charge
is then going to cross the horizon. What happens next is most
interesting.

\subsection{The Blandford-Znajek phase}

The rotational collapse will naturally form a thick equatorial
disk of ionized material around the central black hole. That
material will hold the magnetic flux $\Psi_o$ advected initially
through the horizon and will prevent it from escaping to infinity.
In that phase of the system's evolution, the black hole will spin
down very dramatically according to the Blandford-Znajek
prescription
\begin{equation}
\dot{E} \sim -\frac{1}{6\pi^2 c}\Psi_o^2\Omega^2\  \label{EdotIa}
\end{equation}
(Blandford \& Znajek~1977; Tchekhovskoy, Narayan \& McKinney~2010;
Contopoulos, Kazanas \& Papadopoulos~2013). Here,
\begin{equation}
\Omega=\Omega_o \frac{\alpha}{1+\sqrt{1-\alpha^2}}\ ,
\label{Omega}
\end{equation}
$r_h={\cal G}M(1+\sqrt{1-\alpha^2})/c^2$ and $0\leq \alpha \leq 1$
are the black hole angular velocity, the horizon radius and the
spin parameter respectively.
\begin{equation}
\Omega_o\equiv \frac{c^3}{2{\cal G} M}=10^4\ \ {\rm
rad}/\mbox{sec}
\end{equation}
is the angular velocity of a maximally rotating black hole, and
${\cal G}$ is the gravitational constant. As in pulsars, the
radiated energy is extracted from the available (reducible) black
hole `rotational' energy $\alpha {\cal G}M^2 \Omega/c$
(Christodoulou \& Ruffini~1971). The black hole will therefore
{\em spin down} as
\begin{equation}
\dot{E} = \frac{{\cal G}M^2}{c}\frac{{\rm d}(\alpha\Omega)}{{\rm
d}t}\ . \label{EdotIb}
\end{equation}
We can reverse eq.~(\ref{Omega}) to obtain $\alpha$ as a function
of $\Omega/\Omega_o$ and rewrite the equation that describes the
black hole spindown as
\begin{equation}
\tau_{\rm BZ}\frac{\rm d}{{\rm
d}t}\left(\frac{2(\Omega/\Omega_o)^2}{1+(\Omega/\Omega_o)^2}\right)
=-\left(\frac{\Omega}{\Omega_o}\right)^2\ , \label{EdotIc}
\end{equation}
where
\begin{equation}
\tau_{\rm BZ}\equiv \frac{12 c^5}{{\cal G}^2 B_o^2 M}=33
B_{o16}^{-2} M_{10}^{-1}\ \ \mbox{sec}\ .
\end{equation}
As we will see in the next section, the decay time $\tau_{\rm BZ}$
is {\em a very important physical parameter}. We have defined here
a typical value for the initial black hole magnetic field
\begin{equation}
B_o =\frac{\Psi_o}{\pi r_{ho}^2}=\frac{\Psi_o c^4}{\pi {\cal G}^2
M^2}\ .
\end{equation}
$B_{o16}$ is $B_o$ in units of $10^{16}G$, and $M_{10}$ is the
black hole mass in units of $10M_\odot$.

It is reasonable to assume that, when the black hole forms, it is
maximally rotating. This allows us to integrate eq.~(\ref{EdotIc})
as
\begin{equation}
\frac{1}{1+(\Omega/\Omega_o)^2}+\ln\left(\frac{2(\Omega/\Omega_o)^2}
{1+(\Omega/\Omega_o)^2}\right)=\frac{1-(t/\tau_{\rm BZ})}{2}\ .
\label{EdotId}
\end{equation}
We can solve eq.~(\ref{EdotId}) numerically to obtain
$\Omega=\Omega(t)$, and thus
\begin{equation}
\dot{E}= \frac{-\dot{E}_o}{1+\left[W\left(-\frac{1}{2}{\rm
e}^{-\frac{1+t/\tau_{\rm BZ}}{2}}\right)\right]^{-1}}\ ,
\label{ProductLog}
\end{equation}
where
\begin{equation}
\dot{E}_o\equiv -\frac{\Psi_o^2 \Omega_o^2}{6\pi^2 c}= -3\times
10^{53}B_{o16}^2 M_{10}^2\ \ \mbox{erg}/\mbox{sec}\ .
\label{Edoto}
\end{equation}
$W(x)$ is the Lambert W function which solves the equation
$x=W(x){\rm e}^{W(x)}$. Note that, for a fixed black hole mass,
$\dot{E}_o$ is {\em inversely proportional to $\tau_{\rm BZ}$}. An
approximation to eq.~(\ref{ProductLog}) is
\begin{equation}
\dot{E}\approx \dot{E}_o \frac{{\rm e}^{-t/2\tau_{\rm BZ}}}{2-{\rm
e}^{-t/2\tau_{\rm BZ}}}\ .
\end{equation}

We would like to emphasize that during that phase, the equatorial
disk surrounding the black hole keeps the advected flux in place,
and the black hole magnetic field does not diminish.

\subsection{The pulsar-like phase}

During the Blandford-Znajek phase, the accumulated black hole
electric charge may be estimated by the Wald value
\begin{equation}
Q\sim B_o r_h^2 \frac{\Omega r_h}{c} \label{Q2}
\end{equation}
(Wald~1974). Obviously, this phase will not last for too long.
After a transition period that may last anywhere between a few
minutes to a few weeks, the surrounding material will either be
dispersed away, either be engulfed by the black hole. The black
hole will still be spinning, but it is not clear how much charge
will be left to it, so we can only estimate it through
eq.~(\ref{Q2}).

Let's assume for the moment that all external charges and currents
are removed. An isolated charged and spinning black hole is known
as Kerr-Newman. It is not too well appreciated in the relativity
community that the Kerr-Newman solution {\em is not} a vacuum
solution of the Einstein equations, but a solution that describes
a so called {\em electro-vacuum} with a nonzero electromagnetic
field. There is nothing strange about this result. When the
external currents are removed, a dipolar magnetic field remains
generated by the spinning charge of the black hole (e.g.
Lopez~1983).

The four-potential of the Kerr-Newman electromagnetic field along
the equator is given by $A_\phi=Q\alpha /r$ (e.g.  Misner, Thorne
\& Wheeler~1973; Poisson~2004). It is then straightforward to
calculate the magnetic flux $\Psi_{\rm KN}$ that threads the
horizon as
\begin{equation}
\Psi_{\rm KN} \equiv \int_0^{2\pi}A_\phi(r_h)\frac{2{\cal
G}M}{c^2} {\rm d}\phi = 2\pi Q\alpha\frac{2{\cal G}M}{r_h c^2}\ .
\end{equation}
For a slowly rotating black hole,
\begin{equation}
\Psi_{\rm KN} \approx 2\pi Q \left(\frac{\Omega r_h}{c}\right)
\label{BKN}
\end{equation}
(eq.~\ref{Omega})\footnote{If we estimate $Q$ through
eq.~(\ref{Q2}), our present result differs from Lyutikov \&
McKinney~2011 and Lyutikov~2011 by one extra factor for $(\Omega
r_h/c)$.}. The reader can check that a `maximally charged' slowly
rotating Kerr-Newman black hole corresponds to $Q_{\rm max}\lsim
{\cal G}^{1/2}M$, i.e. to $B_{\rm max}\sim 10^{18}M_{10}^{-2}G$
(eq.~\ref{Q2}), hence values of $B_o\lsim 10^{16}$~G justify the
use of the Kerr metric as an excellent approximation to the
Kerr-Newman one.
The point we would like to emphasize here is that a stellar mass
black hole is very naturally charged during its formation in the
collapse of its progenitor star, and therefore it can naturally
generate its own dipole magnetic field, even after the external
currents are removed.

The astrophysical problem is more complicated. Obviously, the
electromagnetic field cannot remain that of the electro-vacuum
Kerr-Newman solution. Microphysical processes will generate a
distribution of electron-positron pair plasma charges and currents
that will shorten out the electric field component parallel to the
magnetic field. The black hole will absorb opposite charges and
reduce its charge. This effect will be balanced by an equivalent
increase of the rotating magnetospheric charge which is naturally
expected to support an amount of dipolar magnetic flux given
approximately in eq.~(\ref{BKN}). In this picture, the source of
the exterior magnetic field has moved from inside the event
horizon (the Kerr-Newman solution) to just outside
(Petterson~1975; Takahashi \& Koyama~2009). We must acknowledge
that we don't know anything about the stability of such a
configuration besides the fact that if the black hole engulfs the
above magnetospheric charge, it will revert to the Kerr-Newman
solution, so the whole process will start all over again.
Moreover, matching the above exterior solution to an interior
black hole solution is a problem of considerable astrophysical
importance (Ghosh~2000).

During that later {\em pulsar-like} phase of the core collapse,
the spindown of the isolated magnetized black hole will proceed in
analogy to the spindown of the axisymmetric pulsar (e.g.
Punsly~1998). Notice that this is an electrodynamic (not static)
system that holds a rotating magnetospheric electric charge which
we can only assume to decrease as
\begin{equation}
Q\sim B_o r_h^2 \left(\frac{\Omega r_h}{c}\right)^n\ , \label{Q3}
\end{equation}
with $n\geq 0$.

The black hole is not maximally rotating anymore. The
magnetosphere will consist of closed and open field lines, and
only the open field lines (those that cross the light cylinder)
will contribute to the black hole spindown (see Figure~1). Notice
that now $\alpha\ll 1$, therefore, $\Omega=\alpha \Omega_o/2$,
$r_h=2{\cal G}M/c^2$, and the reducible black hole `rotational'
energy is equal to $M (r_h \Omega)^2$ to an excellent
approximation. The axisymmetric pulsar theory now yields
\begin{eqnarray}
\dot{E} & = & 2M r_h^2 \Omega\dot{\Omega} = -\frac{\Psi_{\rm
KN\ open}^2\Omega^2}{6\pi^2 c}\nonumber\\
& = & -B^2 r_h^2 c \left(\frac{\Omega r_h}{c}\right)^4 \sim -Q^2
r_h^{-2} c \left(\frac{\Omega r_h}{c}\right)^6 \ , \label{Edot2}
\end{eqnarray}
which is proportional to $\Omega^{6+2n}$. As before, one can solve
eq.~(\ref{Edot2}) to obtain $\Omega=\Omega(t)$ and
$\dot{E}=\dot{E}(t)$ during this later phase of the black hole
electrodynamic evolution, and show that
\begin{eqnarray}
\Omega & \propto & t^{-1/(4+2n)}\ ,\ \ \mbox{and}\\
\dot{E} & \propto & t^{-(3+n)/(2+n)}\ , \label{KNag}
\end{eqnarray}
Notice that this result is different from the canonical pulsar
spindown (eqs.~\ref{Omegadotp} \& \ref{Edotp}) because a `live'
(astrophysical) pulsar-like black hole loses charge and induced
magnetic flux according to eqs.~(\ref{BKN}) \& (\ref{Q3}). It is
interesting that the power law decay exponent observed during the
GRB afterglow phase has a value between -1 and -1.5 (Nousek {\em
et al.}~2006), in agreement with eq.~(\ref{KNag}). This
observation leads us to associate the pulsar-like phase
(eq.~\ref{KNag}) with the GRB afterglow. Eventually, the black
hole will stop spinning down electrodynamically when its
magnetosphere stops producing the electron-positrons pairs
required to satisfy the force-free condition everywhere, in
analogy to pulsar `death'.

\begin{figure}[t]
\centering
\begin{minipage}[b]{0.5\linewidth}
\includegraphics[trim=0cm 0cm 0cm 0cm,
clip=true, width=6cm, angle=0]{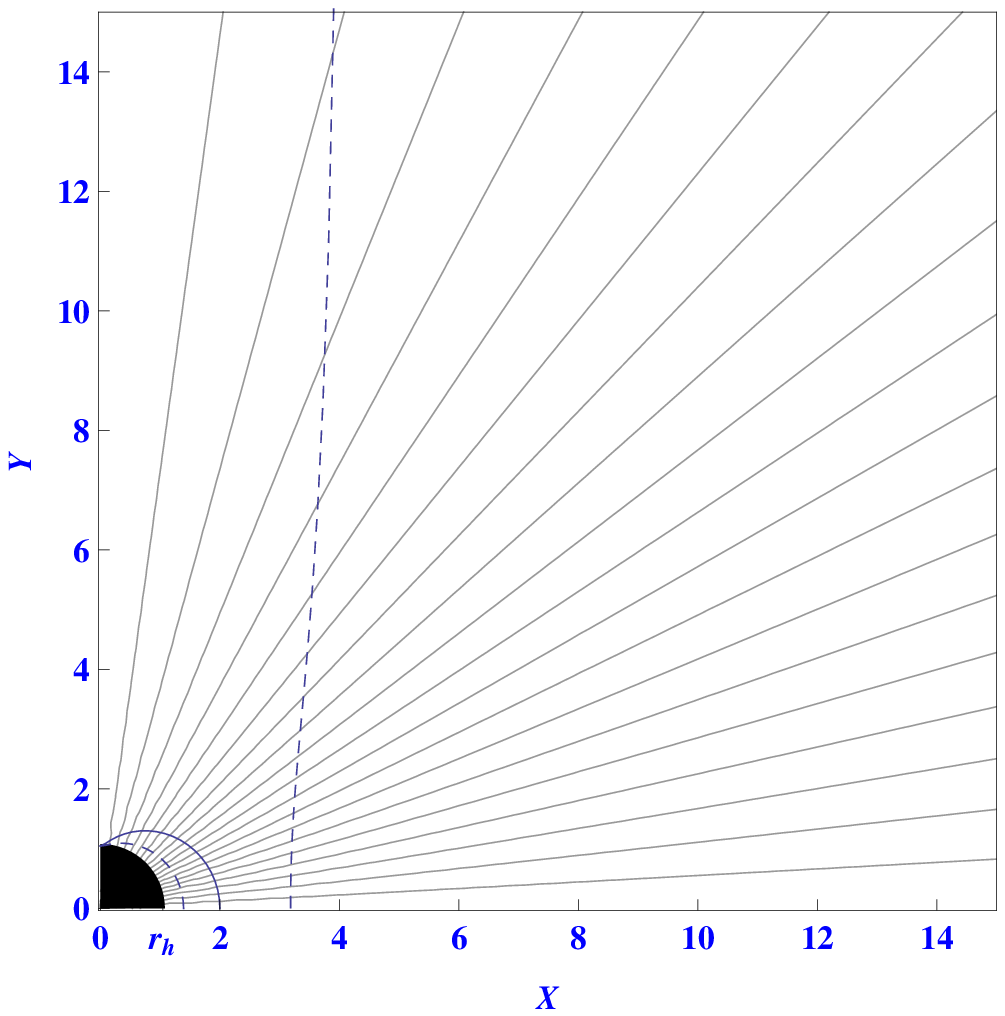}
\includegraphics[trim=0cm 0cm 0cm 0cm,
clip=true, width=6cm, angle=0]{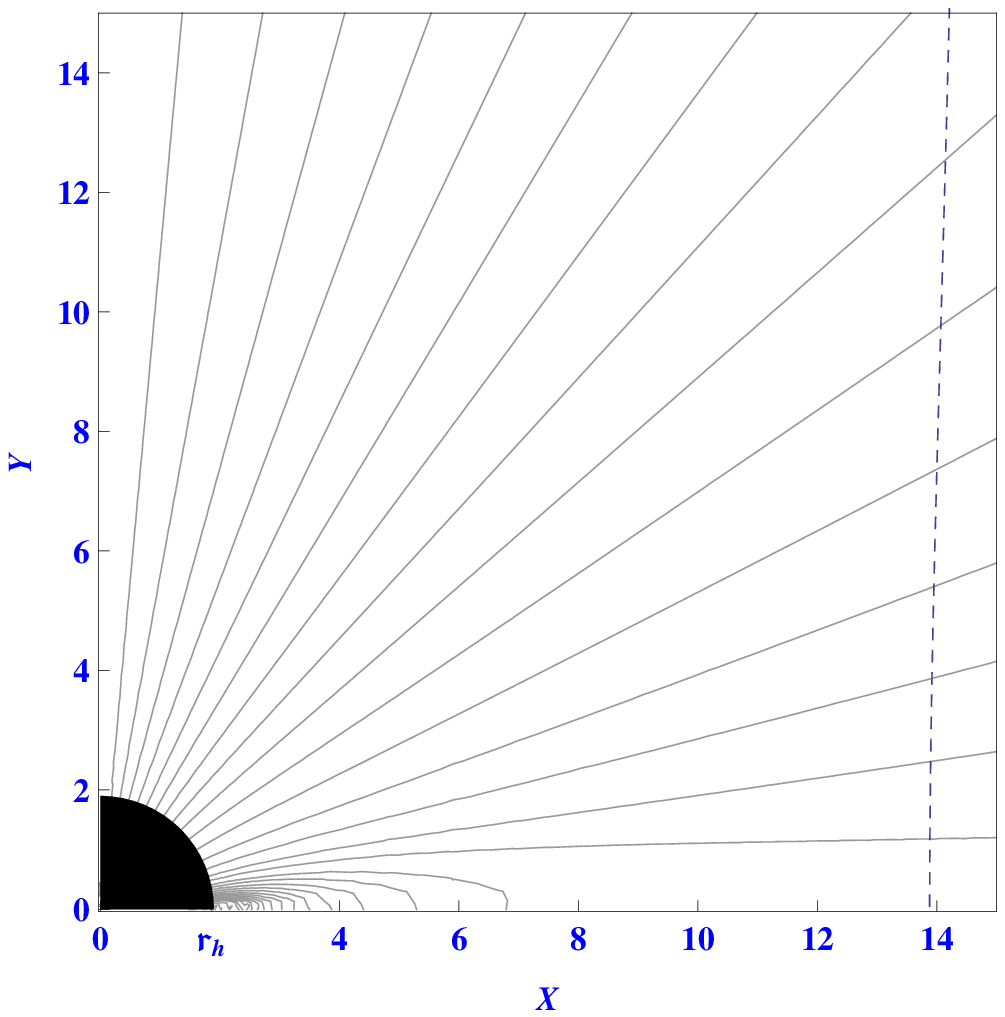}
\end{minipage}
\caption{Poloidal magnetic field lines near the black hole
horizon. To panel: initial Blandford-Znajek phase when the black
hole is maximally rotating and the magnetic flux that threads the
horizon is held in place by the surrounding equatorial material
(Contopoulos, Kazanas \& Papadopoulos~2013). Bottom panel:
pulsar-like phase when the black hole has slowed down by a factor
of about 4 (Pugliese, Contopoulos \& Nathanail 2013, in
preparation). Thicker lines: ergosphere. Dashed lines: light
cylinder and inner light surface.}
\end{figure}

\section{GRB observations}

Our present GRB model of a $10 M_\odot$ newly formed maximally
rotating black hole spinning down electrodynamically explores the
analogy with the axisymmetric pulsar. It is interesting that {\em
neither system} forms relativistic jets on its own, except of
course if there is a surrounding medium that collimates the black
hole/pulsar wind which is nearly isotropic beyond the light
cylinder (Figure~1). As in pulsars, high energy radiation is
generated through reconnection and particle acceleration processes
in the equatorial magnetospheric current sheet (Lyubarsky \&
Kirk~2001; Li, Spitkovsky \& Tchekhovskoy~2012; Kalapotharakos
{\em et al.} 2012). In that respect, our model is {\em
`orthogonal'} to the standard GRB model where all the action takes
place along a relativistic jet emitted along the rotation and
magnetic axis.

We can compare our model with observations. In order to do that,
we need to take into account the source's cosmological redshift
$z$. The observed decay time $\tau_{\rm BZ\ obs}$ is related to
$\tau_{\rm BZ}$ as
\begin{equation}
\tau_{\rm BZ\ obs}=\tau_{\rm BZ} (1+z) \label{C2}\ .
\end{equation}
Straightforward fits of typical GRB light curves (Evans {\em et
al.} 2007, 2009) with initial exponential decay and known
redshifts yield
\begin{equation}
\tau_{\rm BZ} \sim 10-100\ \ \mbox{sec} \label{tauobs}
\end{equation}
(Table~1, Figure~2). Our model predicts that
\begin{equation}
\dot{E}_o \tau_{\rm BZ} =10^{55} M_{10}\ \ \mbox{erg}, \label{erg}
\end{equation}
and therefore, eq.~(\ref{tauobs}) yields $\dot{E}_o \sim
10^{53}-10^{54}$~erg/sec, and $B_{o16}\sim 1$ (eq.~\ref{Edoto}).
Notice that the black hole spin down time is much longer than the
initial rotational period of about 1~msec.

The black hole spindown luminosity $\dot{E}_o$ is not directly
observable. In analogy to pulsars, though, some fraction of it $f$
will be emitted in the form of high energy radiation (X-rays,
$\gamma$-rays) generated by electrons/positrons accelerated
electrostatically in the equatorial magnetospheric current sheet.
For a given luminosity distance ${\rm d}_{\rm L}$ and observed
high energy radiation flux $F$,
\begin{equation}
\dot{E}_{\rm rad}\approx f\dot{E}_o=4\pi f{\rm d}_{\rm L}^2 F
\label{C1}
\end{equation}
under the assumption of isotropic emission. Therefore, in order to
test eq.~(\ref{erg}), one needs to know $f$. If a correlation
between $\tau_{\rm BZ}$ and $\dot{E}_{\rm rad}$ is confirmed in
the few GRB cases with known redshift and a clear exponential
luminosity decay during the initial phase of the burst, this will
allow us to use GRBs as {\em standard candles} in Cosmology.

We conclude that the light curves of long duration gamma-ray
bursts may yield important information about the electrodynamic
processes that take place on the horizon of a spinning black hole.

\acknowledgements{

We would like to thank the referee, Dr. Maxim Lyutikov, whose
sharp criticism led us to reconsider our model for the GRB
afterglow phase. This work made use of data supplied by the UK
Swift Science Data Centre at the University of Leicester, and was
supported by the General Secretariat for Research and Technology
of Greece and the European Social Fund in the framework of Action
`Excellence'.}

\begin{figure}[t]
\centering
\includegraphics[trim=0cm 0cm 0cm 0cm,
clip=true, width=8cm, angle=0]{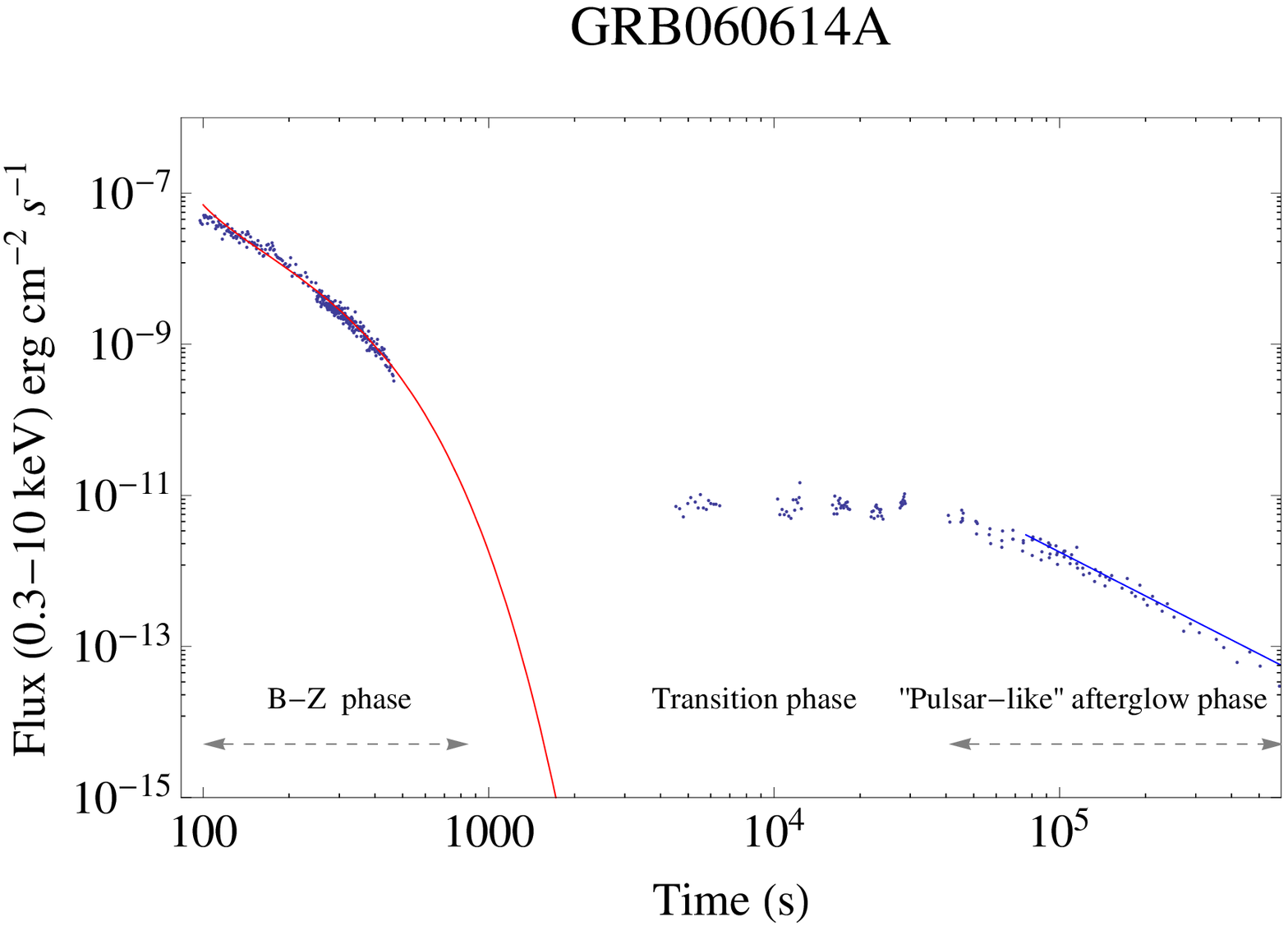} \caption{X-ray light
curve data for GRB 060614A (dots) and fits of early
Blandford-Znajek phase (red line) and late pulsar-like afterglow
(blue line).}
\end{figure}

\begin{deluxetable}{lccccr}
\tablecolumns{7} \tablewidth{0pc} \tablecaption{GRB
Observations$^{\rm a}$}\footnote{Estimates of $F$ from peak
15-150~keV photon flux (Swift data archive)} \tablehead{
\colhead{\em Name} & \colhead{\em z} & \colhead{${\rm d}_{\rm L}$}
& \colhead{$F$} &
\colhead{$\tau_{\rm BZ\ obs}$}  & \colhead{$\dot{E}_{\rm rad}\cdot \tau_{\rm BZ}$}\\ \\
\colhead{} & \colhead{} & \colhead{Gpc} &
\colhead{$\frac{\mbox{erg}}{\mbox{s}\ \mbox{cm}^2}$} & \colhead{s}
& \colhead{$10^{53}$ erg} } \startdata 050502B\footnote{Redshift
estimate from Afonso {\em et al.}~2011} & 5.2 & 50.2 &
$10^{-7}$ & 74    & $4$   \\
060614A & 0.125 & 0.6 & $10^{-6}$ & 48  &  $0.02$\\
080307 &  &  & $10^{-7}$ & 200   & \\
090814A & 2.2 & 17.8 & $10^{-7}$ & 85  &  $1$\\
120401A &  & & $10^{-8}$ & 150   & \\
130701A & 1.155 & 8 & $10^{-6}$ & 180  &  $6$
\enddata
\end{deluxetable}

{}


\begin{thebibliography}{}
\bibitem{} Afonso, B. {\em et al.} 2011, A \& A, 526, 154
\bibitem{} Blandford, R. D. \& Znajek, R. L. 1977, MNRAS, 179,
433
\bibitem{} Christodoulou, D. \& Ruffini, R. 1071, Phys. Rev.
D, 4, 3552
\bibitem{} Contopoulos, I. 2005, A \& A, 442, 579
\bibitem{} Contopoulos, I., Kazanas, D. \& Fendt, C. 1999, ApJ, 511,
351
\bibitem{} Contopoulos, I., Kazanas, D. \& Papadopoulos, D. B. 2013, ApJ, 765,
113
\bibitem{} Evans, P. A. {\em et al.} 2007, A \& A, 468, 379
\bibitem{} Evans, P. A. {\em et al.} 2009, MNRAS, 397,
1177
\bibitem{}Ghosh, P. 2000, MNRAS, 315, 89
\bibitem{} Goldreich, P. \& Julian, W. H. 1969, ApJ,
157, 869
\bibitem{} Kalapotharakos, C., Harding, A. K., Kazanas, D.
\& Contopoulos, I. 2012, ApJ. 754, 1
\bibitem{} Li, J., Spitkovsky, A. \& Tchekhovskoy, A. 2012,
ApJ, 746, 60
\bibitem{} Lopez, C. A. 1983, Nuovo Cimento, 76, 9
\bibitem{} Lyubarsky, Y. \& Kirk, J. G. 2001, ApJ, 547,
437
\bibitem{} Lyutikov, M. 2011, Phys. Rev. D, 83, 124035
\bibitem{} Lyutikov, M. \& McKinney, J. C. 2011, Phys. Rev. D, 84,
084019
\bibitem{} Misner, C. W., Thorne, K. S. \& Wheeler, J. A. 1973,
Gravitation (San Francisco: W. H. Freeman and Co.)
\bibitem{} Nousek, J. A. {\em et al.} 2006, ApJ, 642, 389
\bibitem{} Petterson, J. A. 1975, Phys. Rev. D, 12, 2218
\bibitem{} Poisson, E. 2004, A relativist's toolkit:
the mathematics of black-hole mechanics (Cambridge, UK: Cambridge
Univ. Press)
\bibitem{} Punsly, B. 1998, ApJ, 498, 640
\bibitem{} Takahashi, M. \& Koyama, H., 2009, ApJ, 693, 472
\bibitem{} Tchekhovskoy, A., Narayan, R. \& McKinney, J. C.
2010, ApJ, 711, 50
\bibitem{} Wald, R. M. 1974, Phys. Rev. D, 10, 1680
\end{thebibliography}
\end{document}